\documentclass[longauth]{aa} 

\usepackage{graphicx}
\usepackage{hyperref}
\usepackage{natbib}
\usepackage{txfonts}

\newcommand{\oca}{1}
\newcommand{\luan}{2}
\newcommand{\laog}{3}
\newcommand{\mpifr}{4}
\newcommand{\oaa}{5}

\begin{document}
\title{An asymmetry detected in the disk of $\kappa$ CMa\thanks{Based on Guaranteed Time Observations made with the Very Large Telescope
    Interferometer at Paranal Observatory} with the VLTI/AMBER
}
\titlerunning{An asymmetry detected in the disk of $\kappa$ CMa}

\institute{Laboratoire Gemini, U.M.R. 6203 Observatoire de la C\^ote d'Azur/C.N.R.S.,
  Avenue Copernic, 06130 Grasse, France
  \and
  Laboratoire Universitaire d'Astrophysique de Nice, UMR 6525
  Universit\'e de Nice/CNRS, Parc Valrose, F-06108 Nice cedex 2, France
  \and
  Laboratoire d'Astrophysique de Grenoble, U.M.R. 5571 Universit\'e
  Joseph Fourier/C.N.R.S., BP 53, F-38041 Grenoble Cedex 9, France
  \and
  \and Max-Planck-Institut f\"ur Radioastronomie, Auf dem H\"ugel 69,
  D-53121 Bonn, Germany
  \and INAF-Osservatorio Astrofisico di Arcetri, Istituto Nazionale di
  Astrofisica, Largo E. Fermi 5, I-50125 Firenze, Italy
  \and ONERA/DOTA, 29 av de la Division Leclerc, BP 72, F-92322
  Chatillon Cedex, France 
  \and Centre de Recherche Astronomique de Lyon, UMR 5574 Universit\'e
  Claude Bernard/CNRS, 9 avenue Charles Andr\'e, F-69561 Saint Genis
  Laval cedex, France
  \and Division Technique INSU/CNRS UPS 855, 1 place Aristide
  Briand, F-92195 Meudon cedex, France
  \and IRCOM, UMR 6615 Universit\'e de Limoges/CNRS, 123 avenue Albert
  Thomas, F-87060 Limoges cedex, France
  \and European Southern Observatory, Karl Schwarzschild Strasse 2,
  D-85748 Garching, Germany
  \and European Southern Observatory, Casilla 19001, Santiago 19,
  Chile
  \and Instituut voor Sterrenkunde, KULeuven, Celestijnenlaan 200B,
  B-3001 Leuven, Belgium 
  \and Centro  de  Astrof\'{\i}sica  da  Universidade  do  Porto, Rua
  das Estrelas - 4150-762 Porto, Portugal 
  \and \emph{Present affiliation:} Observatoire de la Côte d'Azur -
  Calern, 2130 Route de l'Observatoire , F-06460 Caussols, France
  \and \emph{Present affiliation:} Laboratoire Astrophysique de
  Toulouse, UMR 5572 Universit\'e Paul Sabatier/CNRS, BP 826, F-65008
  Tarbes cedex, France 
}

\author{A.~Meilland  \inst{\oca}
  \and  F.~Millour\inst{\luan,\laog}   
  \and Ph.~Stee\inst{\oca}
  \and A.~Domiciano~de~Souza \inst{\luan,\oca}
  \and R.G.~Petrov \inst{\luan}      
  \and D.~Mourard\inst{\oca}   
  \and S.~Jankov \inst{\luan}   
  \and S.~Robbe-Dubois\inst{\luan} 
  \and A.~Spang \inst{\oca}  
  \and E.~Aristidi\inst{\luan}
  \and
  \\P.~Antonelli\inst{\oca}
  \and U.~Beckmann\inst{\mpifr}
  \and Y.~Bresson\inst{\oca}
  \and A.~Chelli\inst{\laog}
  \and M.~Dugu\'e\inst{\oca}
  \and G.~Duvert\inst{\laog}
  \and L.~Gl\"uck\inst{\laog}
  \and P.~Kern\inst{\laog}
  \and S.~Lagarde\inst{\oca}
  \and E.~Le Coarer\inst{\laog}
  \and F.~Lisi\inst{\oaa}
  \and F.~Malbet\inst{\laog}
  \and K.~Perraut\inst{\laog}
  \and P.~Puget\inst{\laog}
  \and G.~Weigelt\inst{\mpifr}
  \and G.~Zins\inst{\laog}
  \and \\
  M.~Accardo\inst{\oaa}
  \and B.~Acke\inst{\laog,12}
  \and K.~Agabi\inst{\luan}
  \and B.~Arezki\inst{\laog}
  \and E.~Altariba\inst{\luan}
  \and C.~Baffa\inst{\oaa}
  \and J.~Behrend\inst{\mpifr}
  \and T.~Bl\"ocker\inst{\mpifr}
  \and S.~Bonhomme\inst{\oca}
  \and S.~Busoni\inst{\oaa}
  \and F.~Cassaing\inst{6}
  \and J.-M.~Clausse\inst{\oca}
  \and C.~Connot\inst{\mpifr}
  \and A.~Delboulb\'e\inst{\laog}
  \and T.~Driebe\inst{\mpifr}
  \and P.~Feautrier\inst{\laog}
  \and D.~Ferruzzi\inst{\oaa}
  \and T.~Forveille\inst{\laog}
  \and E.~Fossat\inst{\luan}
  \and R.~Foy\inst{7}
  \and D.~Fraix-Burnet\inst{\laog}
  \and A.~Gallardo\inst{\laog}
  \and S.~Gennari\inst{\oaa}
  \and A.~Glentzlin\inst{\oca}
  \and E.~Giani\inst{\oaa}
  \and C.~Gil\inst{\laog,13}
  \and M.~Heiden\inst{\mpifr}
  \and M.~Heininger\inst{\mpifr}
  \and O.~Hernandez\inst{\laog} 
  \and K.-H.~Hofmann\inst{\mpifr}
  \and D.~Kamm\inst{\oca}
  \and S.~Kraus\inst{\mpifr}
  \and D.~Le Contel\inst{\oca}
  \and J.-M.~Le Contel\inst{\oca}
  \and B.~Lopez\inst{\oca}
  \and Y.~Magnard\inst{\laog}
  \and A.~Marconi\inst{\oaa}
  \and G.~Mars\inst{\oca}
  \and G.~Martinot-Lagarde\inst{8,14}
  \and P.~Mathias\inst{\oca}
  \and J.-L.~Monin\inst{\laog}
  \and D.~Mouillet\inst{\laog,15}
  \and P.~M\`ege\inst{\laog}
  \and E.~Nussbaum\inst{\mpifr}
  \and K.~Ohnaka\inst{\mpifr}
  \and F.~Pacini\inst{\oaa}
  \and C.~Perrier\inst{\laog}
  \and Y.~Rabbia\inst{\oca}
  \and S.~Rebattu\inst{\oca}
  \and F.~Reynaud\inst{9}
  \and A.~Richichi\inst{10}
  \and A.~Roussel\inst{\oca}
  \and M.~Sacchettini\inst{\laog}
  \and P.~Salinari\inst{\oaa}
  \and D.~Schertl\inst{\mpifr}
  \and W.~Solscheid\inst{\mpifr}
  \and P.~Stefanini\inst{\oaa}
  \and M.~Tallon\inst{7}
  \and I.~Tallon-Bosc\inst{7}
  \and D.~Tasso\inst{\oca}
  \and E.~Tatulli\inst{\laog}
  \and L.~Testi\inst{\oaa}
  \and J.-C.~Valtier\inst{\oca}
  \and M.~Vannier\inst{\luan,11}
  \and N.~Ventura\inst{\laog}
  \and M.~Kiekebusch\inst{11}
  \and M.~Sch\"oller\inst{11}
}

\offprints{anthony.meilland@obs-azur.fr}

\date{Received; accepted }

\abstract{}
{We study the geometry and kinematics of the circumstellar environment of the Be star $\kappa$ CMa in the Br$\gamma$ emission line and its nearby continuum.}
{We use the VLTI/AMBER instrument operating in the K band which
  provides a spatial resolution of about 6 mas with a spectral
  resolution of 1500 to study the kinematics within the disk and to
  infer its rotation law. In order to obtain more kinematical
  constraints we also use an high spectral resolution Pa$\beta$ line
  profile obtain in December 2005 at the Observatorio do Pico do Dios,
  Brazil and we compile V/R line profile variations and spectral
  energy distribution data points from the literature.} 
{Using differential visibilities and differential phases across the
  Br$\gamma$ line we detect an asymmetry in the disk. Moreover,
 we found that $\kappa$ CMa seems difficult to fit within the classical scenario for
  Be stars,  illustrated recently by $\alpha$ Arae observations,
  i.e. a  fast rotating B star close to its breakup velocity surrounded by a 
  Keplerian circumstellar disk with an enhanced polar wind. Finally we
 discuss the possibility for $\kappa$ CMa to be a critical rotator with
 a Keplerian rotating disk and try to see if the detected asymmetry can be 
 interpreted within  the "one-armed" viscous disk framework.}
{}

\keywords{   Techniques: high angular resolution --
  Techniques: interferometric  --
  Stars: emission-line, Be  --
  Stars: Keplerian rotation --
  Stars: individual ($\kappa$ CMa) --
  Stars: circumstellar matter
}

\maketitle
%

\section{Introduction}
The "Be phenomenon" is related to hot stars that have at least
exhibited once Balmer lines in emission with infrared excess produced
by free-free and free-bound processes in an extended circumstellar
disk. There is now a strong evidence that the disk around the
Be star $\alpha$ Arae is Keplerian (Meilland et
  al.~\citealp{meillanda}) and that this dense
equatorial disk is slowly expanding. On the other side there are also
clear pieces of evidence for a polar enhanced wind. This was already predicted
for almost critically rotating stars as for a large fraction of Be
stars.  Recently, Kervella \& Domiciano de Souza~\cite{kervella}
showed an enhanced polar wind for the Be star Achernar whereas this Be
star presents no hydrogen lines in strong emission. Thus, it seems
that a significant polar wind may be present even if the star is still
in a normal B phase, signifying this enhanced polar wind would not be
related to the existence of a dense equatorial envelope.  However many
issues remain unsolved on the actual structure of the circumstellar
envelopes in Be stars which probably depends on the dominant mass
ejection mechanisms from the central star and on the way the ejected
mass is redistributed in the near circumstellar environment. Recently
Meilland et al.~\cite{meillandb} reported theoretical spectral energy
distributions (SEDs), Br$\gamma$ line profiles and visibilities for
two likely scenarii of the disk dissipation of active hot stars, and
account for the transition from the Be to the B spectroscopic
phase. \\
$\kappa$ CMa (HD 50013, HR 2538) is one the brightest Be star of the
southern hemisphere (V=3.8, K=3.6). It is classified as a B2IVe star,
and the distance deduced from Hipparcos parallax is 230 $\pm$30
pc. The measured vsini values range from  220 kms$^{-1}$ (Dachs et
  al.~\citealp{dachs0}; Mennickent et al.~\citealp{mennickent};
  Okazaki~\citealp{okazaki}; Prinja~\citealp{prinja}) to 243 kms$^{-1}$ (Zorec 
  et al.\citealp{zorec}) , its radius is 6
  R$_{\sun}$ (Dachs et al.~\citealp{dachs0}; Prinja~\citealp{prinja})
  and its mass is 10 M$_{\sun}$ (Prinja~\citealp{prinja}).\\
  
  \noindent We must mention that the mass and radius determination of a Be star is not
  an easy task. For instance if we assume values of masses and radii from Harmanec~\cite{harmanec1} compilation,
  in agreement with Schaller et al.~\cite{schaller} non-rotating evolutionary models, for the effective temperatures used by Popper~\cite{popper}, Prinja~\cite{prinja} and Fremat~\cite{fremat}  we obtain the table~\ref{tableradius}.\\
    
\begin{table}[htbp]
  \begin{center}
  \begin{tabular}{cccc} \hline
T$_{eff}$  & Mass & Normal Radius & Radius from \\
in K		&    in M$_{\odot}$      &   in R$_{\odot}$   &parallax in R$_{\odot}$\\
    \hline \\
20000   &   6.60   &       3.71   &       7.25  (6.46 - 8.24)\\
23100   &   8.62   &       4.28    &     6.26  (5.59 - 7.13)\\
25800   & 10.72  &        4.83   &     5.59  (4.98 - 6.36)\\
    \hline
    \end{tabular}
  \end{center}  
    \caption{Masses and radii determination for $\kappa$ CMa from Harmanec~\cite{harmanec1} compilation for the effective temperatures given by Popper~\cite{popper}, Prinja~\cite{prinja} and Fremat~\cite{fremat}.}
      \label{tableradius}
\end{table}
  
 \noindent Thus, for a main sequence star the stellar radius should be smaller than the 6 R$_\odot$ we
 have adopted but on the other side, our radius estimate based on the parallax and the
 chosen V magnitude from the correlation between the brightness and emission strength, as
 proposed by Harmanec~\cite{harmanec0}, gives the range of radii comparable to our chosen 6 R$_\odot$ used
 in our modeling.

 \noindent The star exhibits a huge IR-excess and a strong emission in the hydrogen lines
making a good candidate for the VLTI/AMBER spectro-interferometer
(Petrov et al. \cite{petrov}) using medium spectral resolution
(1500). Our aim is to study the geometry and kinematics of the
circumstellar environment of this star as a function of wavelength,
especially across the Br$\gamma$ emission line and to detect any
signature of a possible asymmetry of its circumstellar disk as already
observed through a violet to red peaks ratio V/R $\sim$ 1.3 by Dachs
et al.~\cite{dachs} and Slettebak et al.~\cite{Slettebak}.

\section{Observations and data reduction}
Dedicated observations of $\kappa$ CMa were carried out during the
night of the December, 26th 2004 with the three VLTI 8m ESO telescopes
UT2, UT3 and UT4 (See Table \ref{table_visi} for the baseline
configurations). The
data were reduced using the amdlib (v1.15)/ammyorick (v0.54) software
package developed by the AMBER consortium. It uses a new data
processing algorithm adapted to multiaxial recombination instruments
called P2VM for {\it Pixel To Visibility Matrix} algorithm. The
squared visibility estimator is computed from the basic observable
coming from this algorithm that is the coherent flux (i.e. complex
visibilities frame by frame multiplied by the flux) and the estimated
fluxes from each telescope. The principles of the general AMBER data
reduction are described in more details by Millour et
al. \cite{millour1} and Tatulli et al. \cite{tatulli}.

The complex coherent flux allows also to compute differential phase,
i.e. averaged instantaneous phase substracted from achromatic
atmospheric OPD and a wavelength-averaged reference phase. 
This means that the differential phase is the difference between the
phase of the source complex visibility and a mean OPD. This leads to
an average differential phase equal to zero on the observed spectral
window and the lost of the object's phase slope over the
wavelengths. This technique allows however to retrieve partial
information of the object's phase and is almost equal to the object's
interferometric phase if we have some spectral channels where we know
the object's phase is zero.

Moreover, it also allows to compute ``differential'' visibility 
  (as defined in Millour et al. \citealp{millour} and in Meilland et
  al. \citealp{meillanda}), i.e. the instantaneous modulus of the
complex visibility divided by the averaged visibility on all the
wavelengths excepted the work one. This leads to an average differential
visibility equal to 1 in the continuum. It has the advantage over the
``classical'' visibility estimator to be almost insensitive to rapid
frame to frame variations of visibility (due to vibrations or
atmospheric jittering for example) and therefore one can expect the
differential visibility observable to be more precise than the
classical visibility estimator given the current vibrations in the
VLTI infrastructure, and even though the continuum visibility
information is lost in this observable.

For more information, the differential data reduction is described in
details in Millour et al. \cite{millour0} and Millour et
al. \cite{millour}.

Reducing the $\kappa$ CMa data with a good accuracy was quite
difficult to achieve. We encountered specific problems related
to this data set. Therefore, in addition to the tools furnished
by the default package, some specific processing was added to
reach the best precision on the interferometric observables.

\begin{table}[htbp]
  \centering
  \caption{
    \footnotesize{
      Calibration stars diameters estimated from spectro-photometric
      indices (computed as in Bonneau et al. \citealp{Bonneau})
      and their associated errors.
    }
  }
  \label{tableDiam}

  \begin{tabular}{ccc}
    \hline

    Star     & Diameter (mas) & Error (mas) \\

    \hline \\

    HD 75063 & 0.50           & 0.08\\
    HD 93030 & 0.454          & 0.006\\

    \hline
  \end{tabular}
\end{table}

\begin{figure}[htbp]
  \centering
  \includegraphics[width=0.45\textwidth]{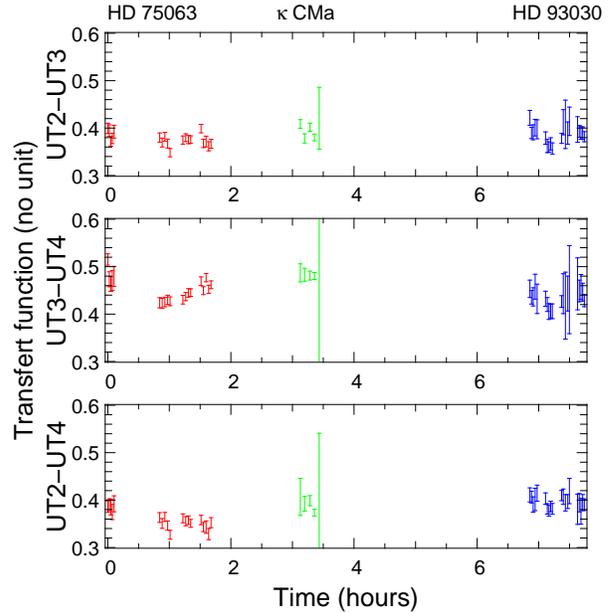}
  \caption{
    \footnotesize{
      Raw absolute visibilities of calibration stars corrected from
      their angular diameters and averaged over the [2.13-2.21] $\mu$m
      window, allowing us to monitor the    instrumental+atmospheric
      transfert function (points respectively around 1h in red and 7h
      in blue). We have overplotted for comparison the raw visibilities of
      $\kappa$ CMa (around 3h in green). The $\kappa$ CMa visibilities
      have obviously the same value as the instrumental+atmospheric
      transfert function within the error bars, leading to a
      calibrated visibility of 1, i.e. a non resolved or very barely
      resolved object on all baselines.
    }
  }
  \label{figure-Vis}
\end{figure}

\begin{itemize}
\item First of all, no specific data were available to calibrate the
  fringe contrast of $\kappa$ CMa. We therefore looked at calibration
  stars observed during the same night for other stars and corrected
  their visibilities averaged over all the [2.13-2.21] $\mu$m observed
  spectral range from their estimated diameters (see table
  \ref{tableDiam}) in order
  to monitor the instrumental+atmospheric transfert function (see
  Fig. \ref{figure-Vis}). This transfert function is the
    visibility of a point source measured by the instrument, allowing
    us to correct the raw visibilities on the science star from the
    instrumental-specific visibility loss. The scattering over the
  time of the visibilities gives the dispersion due to the
  instrumental drifts and atmospheric fluctuations during the
  observing time. This leads to a visibilities dispersion estimate of
  0.05 for each star, which leads to an error on calibrated
  visibilities of 0.07 ($\sqrt{2 \times 0.05^2}$).\\
  Then we interpolate the estimated transfert function to the time of
  the science star observations (as in Perrin et
  al. \citealp{perrin}). By using this technique, we find that the
  [2.13-2.21] $\mu$m averaged visibility of $\kappa$ CMa is really
  close to 1.0 with an uncertainty of 0.07 on all the observed base
  lengths. This would normally be killing for the
  wavelength-dependence study of the visibilities, but as explained
  before, we expect to have differential visibility and differential phase
  estimators much more precise than the visibility estimator.
\item The lack of dedicated calibration star for $\kappa$ CMa should
  have lead to an unfeasibility to spectrally calibrate the
  differential observables, but fortunately an other calibration star
  (HD93030) was observed two hours later within almost the same
  spectral window, which means that the spectrograph grating did not
  move but that the  detector window was not exactly the same as for
  $\kappa$ CMa, allowing us to use the intersecting spectral channels
  between the two  observations without any calibration
  problem. Detailed data analysis of calibration stars tends to
  demonstrate that the main pattern on differential observables comes
  from a fiber-injection pattern (i.e. a pure internal AMBER
  instrumental effect) and that it is stable over several minutes in
  the 10$^{-2}$ range for the differential visibilities and 10$^{-2}$
  radians for the differential phase at medium spectral resolution
  (R $\sim$ 1500, see for instance Vannier et al., \citealp{vannier}).
\end{itemize}

\begin{figure}
  \begin{center}
    \includegraphics[width=7.0cm, height=13.0cm]{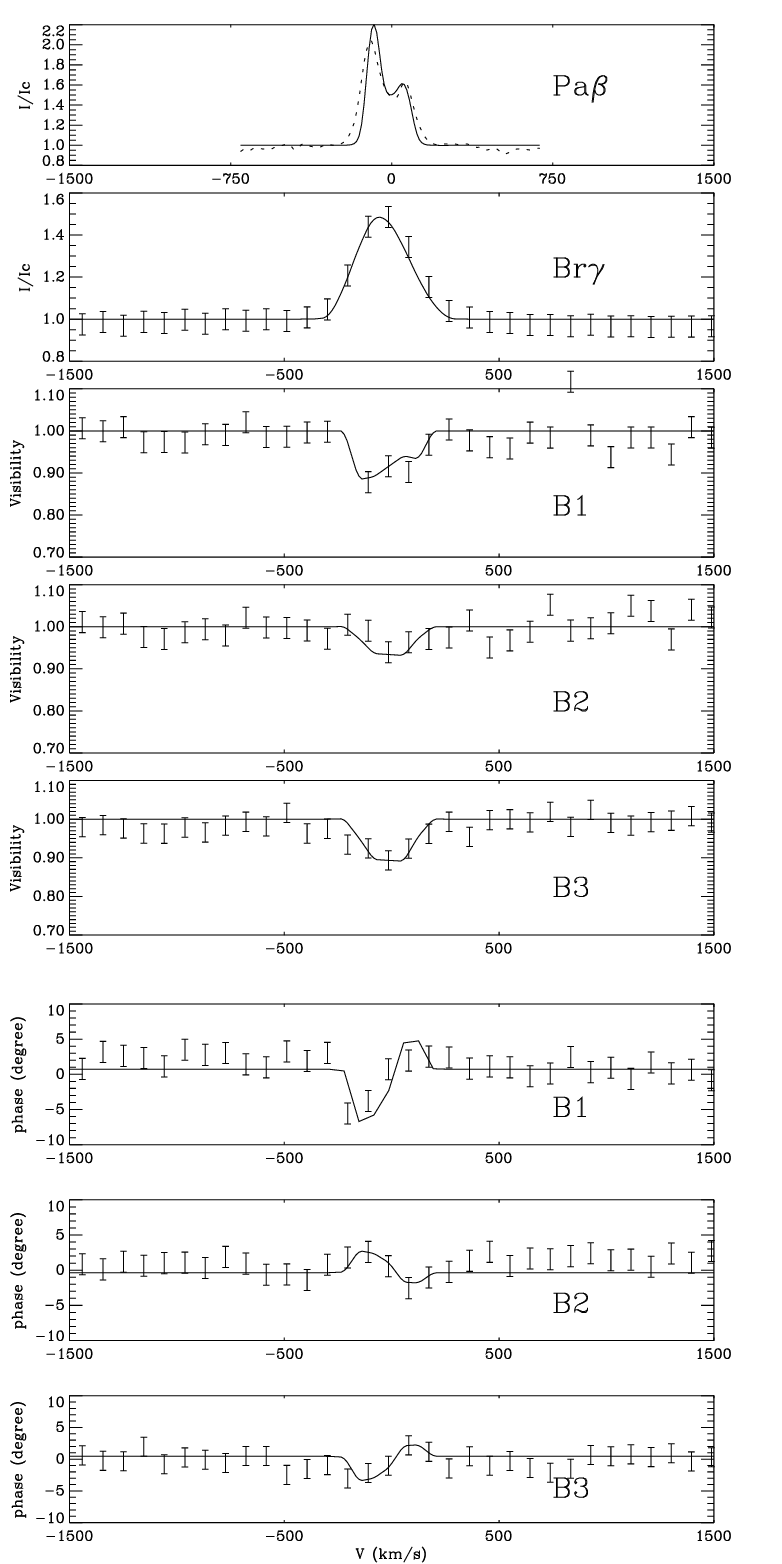}
    \caption{
      \footnotesize{
        From the top to bottom : Pa$\beta$ line profile from the
        Observatorio do Pico dos Dias, Brazil (dotted line) with our
        best model fit (plain line), Br$\gamma$ line profile, differential
        visibilities and differential phases for the three
        baselines. For each plot, the dots with errors bars are
        VLTI/AMBER data and the solid line is from our best SIMECA model
        (see section 4).
      }
    }	
    \label{figure_asymmodel}	
  \end{center}	
\end{figure}

This allowed us to calibrate correctly the differential visibility and
the differential phases (See Fig. \ref{figure_asymmodel}). In order to
ensure our calibration, we checked that all the features mentioned in
this article are already present in the uncalibrated data, and not
added by pure noise-effects produced by the calibrator star.

We could expect to see an effect in the closure phase, but its
modulation seems to be of the order of amplitude of the error
bars (~3 \degr or 0.05 radians), which means that we do not see any
detectable signal in the closure phase. This non-detection confirms
the result on the visibility and the low amplitude of the modulation
on the differential phases: the object is almost non-resolved or
barely resolved by the interferometer on the considered baselines (80m
maximum).

What we see in the observed data is a decrease in the differential
visibilities in two of the three baselines of the order of 0.07, much
larger than the error bars (0.02 for the differential
visibilities). This can be explained by an envelope larger than the
star, visible in the emission line.

We observe also a modulation in the differential phase of
the order of 5\degr (0.09 radians), also higher than the error bars
(~2\degr or 0.03 radians). The modulation of the differential phase
show a ``sine arch'' shape, typical of a rotating object or a bipolar
outflow but also show an asymmetry, mainly on the baseline UT2-UT3
(B1).

In order to obtain more kinematical constraints the star has also
  been observed in the J2 band (1.2283-1.2937$\mu$m) with the
  1.6 m Perkin-Elmer telescope and Coud\'e spectrograph (with R=10\,000) at the
  Observat\'orio do Pico dos Dias, Laborat\'orio Nacional de
  Astrof\'{\i}sica (LNA), Itajub\'a, Brasil. The spectra were recorded
  in the night 20/21 November 2005, at seven different positions along
  the slit using the C\^amara Infravermelho (CamIV) detector. The
  images of the dark have been subtracted from each star's spectral
  image, wavelength calibration image and five flat-field images. To
  get the sky image we made the median combination of the star's
  spectral images (divided previously by the average of flat-field
  images). The sky image has been subtracted from stellar images and
  the one-dimensional spectra were extracted and calibrated in
  wavelength using the standard IRAF \footnote{IRAF is distributed by
    the National Optical Astronomy Observatories, which is operated by
    the Association of Universities for Research in Astronomy (AURA),
    Inc., under cooperative agreement with the National Science
    Foundation.} procedures. Finally the continuum normalization
  around the Pa$\beta$ line has been performed using our software. The
  average profile of the line, which has been used to constrain the
  kinematics within the disc, is plotted in Fig.~\ref{figure_asymmodel} and
  \ref{figure_pabeta}.

\section{Envelope extension and flattening}
In this section we present the AMBER data in order to obtain an
estimate of the $\kappa$ CMa's envelope geometry and
extension. Assuming that the measured visibility in the continuum,
$V_c$, is only due to the central star and its circumstellar envelope
and that the envelope is optically thin in the continuum,  we can
write:

\begin{equation}
  V_c= \frac {V_{ec} F_{ec} + V_{\star} F_{\star}} {F_{c}}
\end{equation}
where V$_{ec}$ and F$_{ec}$ are respectively the envelope visibility
and flux in the continuum, V$_{\star}$ and F$_{\star}$ are the star
visibility and flux in the continuum and F$_{c}$ = F$_{ec}$+F$_{\star}$. 

\begin{figure}[htbp]
  \begin{center}
    \includegraphics[height=6.0cm, width=8cm]{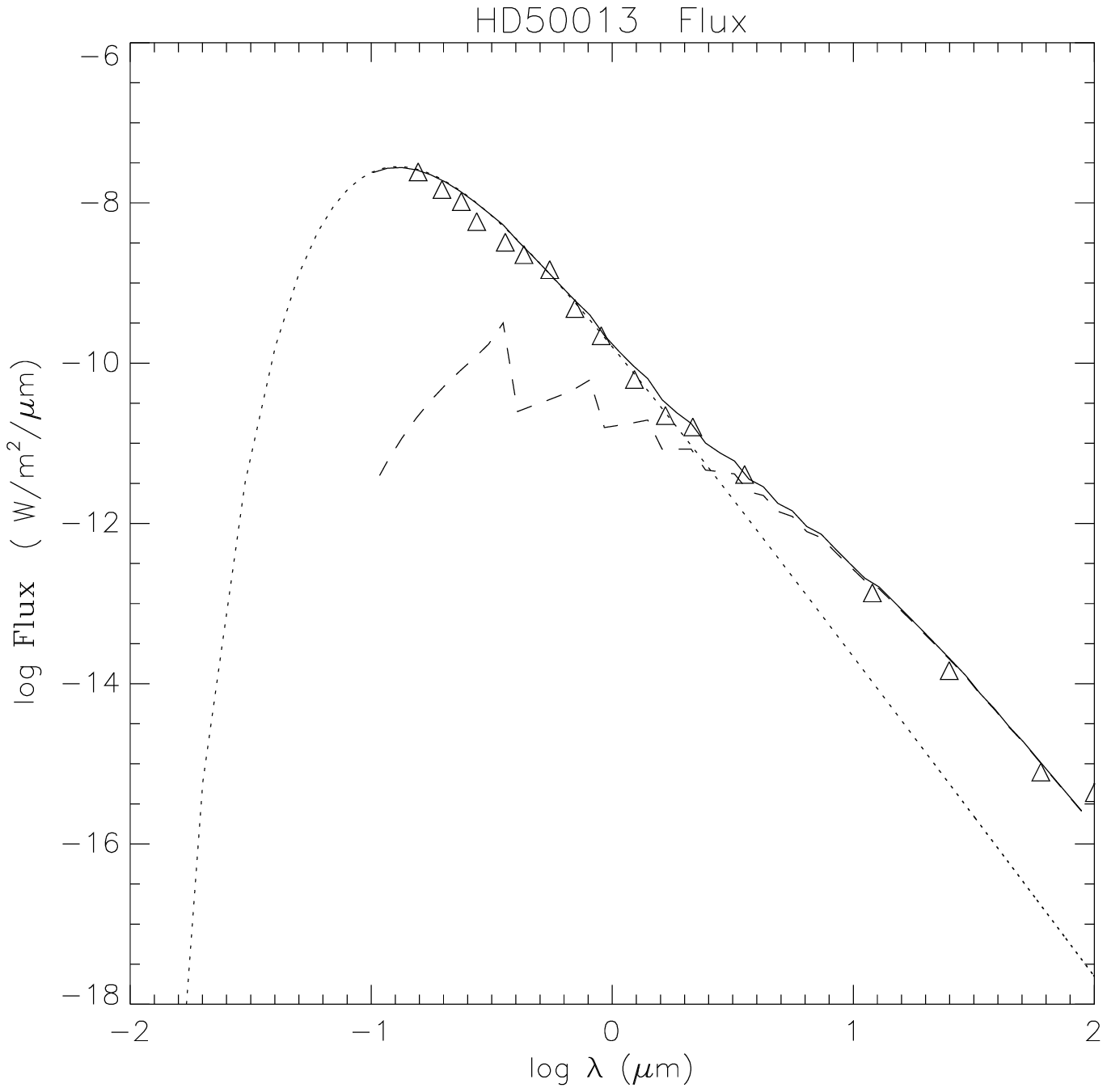}		       
    \caption{
      \footnotesize{
        $\kappa$ CMa's Spectral Energy Distribution (SED) from
        SIMBAD CDS (triangles). Dotted line: emission from the central
        star assuming a black body with $R_\star = 6 R_\odot$,
        T$_{eff}$=22500K and d=230 pc. Dashed line : free-free and
        free-bound envelope contribution from the SIMECA code between
        0.3 and 100 $\mu$m . Plain line : Central star emission +
        envelope contribution.
      }
    }
    \label{figure_sed}
  \end{center}
\end{figure}

The total flux is normalized, i.e. F$_{c}$ = F$_{ec}$ + F$_{\star}$ =
1. Since the star is fully unresolved $\phi_{\star}\sim 0.25$mas
(assuming a 6 R$_{\sun}$ seen at 230 pc) which corresponds to
$V_{\star} > 0.99$ for the longest baseline at 2.1 $\mu$m, we assume
in the following that V$_{\star}$=1. In order to estimate V$_{ec}$ we
still have to determine the star and the envelope contributions at 2.1
$\mu$m.  Using the fit of the SED given in Fig.~\ref{figure_sed} we estimate
that at this wavelength the stellar emission is similar to the
envelope contribution, i.e.  F$_{\star}$ = F$_{ec}$ = 0.5.

We have the same relation for the visibility in the Br$\gamma$ line:
\begin{equation}
  V_{r}= \frac {V_{er} F_{er} + V_c F_c} {F_r}
\end{equation}
where V$_r$ and F$_r$ are respectively the measured visibility and
flux in the Br$\gamma$ line. V$_c$ and F$_{c}$ are previously defined
and V$_{er}$ and F$_{er}$ are the visibility and flux only due to the
Br$\gamma$ line, i.e. without any stellar contribution and envelope
continuum. Using the AMBER Br$\gamma$ emission line profile plotted in
Fig.~\ref{figure_asymmodel} and neglecting the underlying broadened
photospheric absorption line, we obtain F$_{er}$=0.5 and F$_r$=1.5 at
the center of the line.

The  corresponding visibilities, deduced from Eq. 1 and 2 and from the
measurements shown in Fig.~\ref{figure_asymmodel}, are given in Table
\ref{table_visi}. Using a uniform disk model for the envelope
contribution, for each measurement,  we also estimate in Table
\ref{table_visi}  the corresponding angular diameters in the continuum
and in the Br$\gamma$ line. Since the envelope is marginally resolved
in the continuum we simply put an upper limit to its angular size.

\begin{figure}[htbp]
  \begin{center}
    \includegraphics[height=8.0cm]{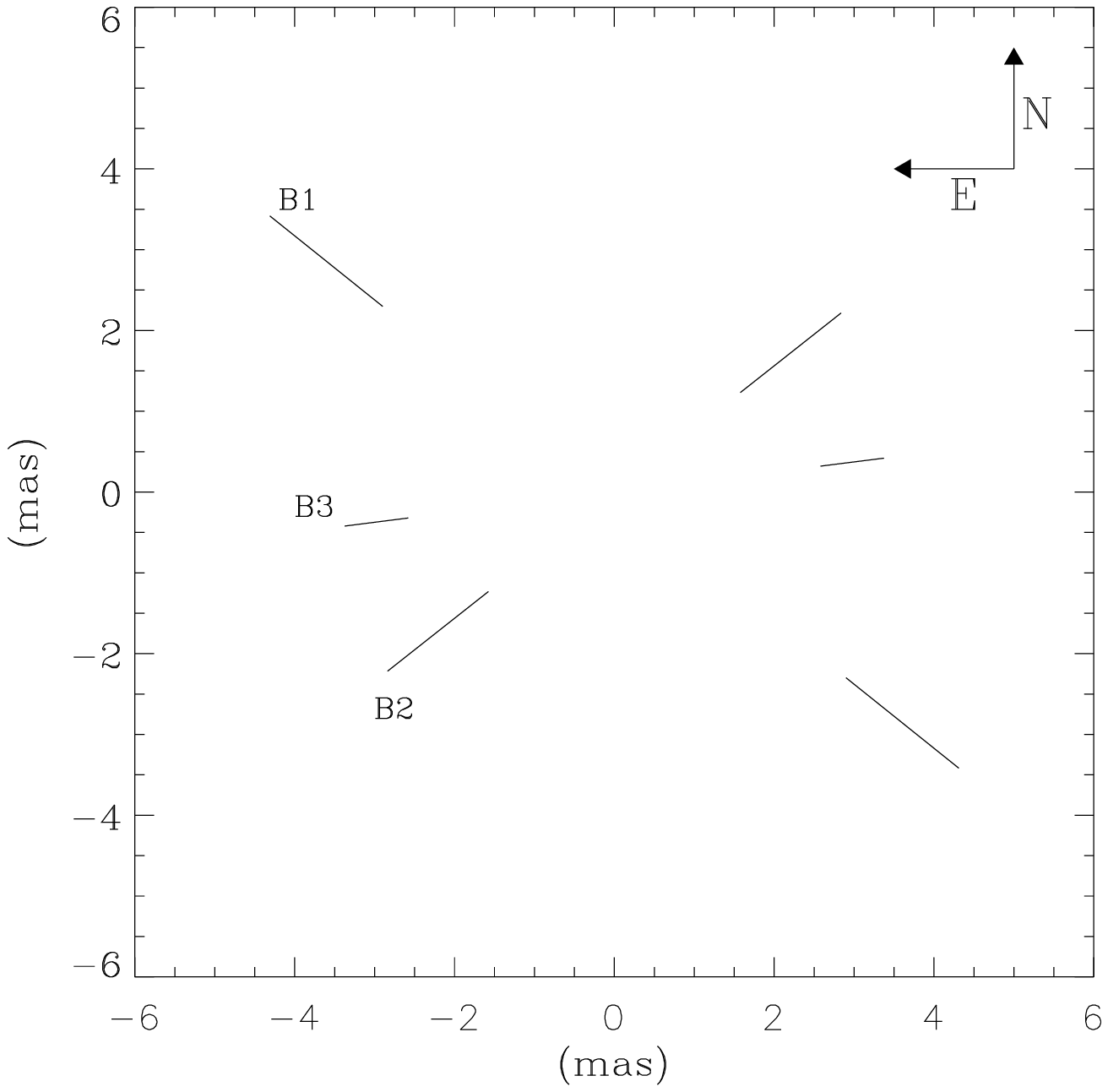}
    \caption{
      \footnotesize{
        $\kappa$ CMa diameters in the Br$\gamma$ line, assuming
        an unresolved star + uniform disk models, as a function of the
        baseline position angle (in mas). The length of each plot
        corresponds to the error bar measurement whereas diameters are
        given by the center of each error bar.
      }
    }
    \label{figure_elong_env}	
  \end{center}
\end{figure}

The envelope extensions in Br$\gamma$ given in Table \ref{table_visi}
are strongly dependent on the sky-plane baseline orientation as seen
in Fig.~\ref{figure_elong_env}, where we plotted the $\kappa$ CMa
(unresolved star + uniform disk) model diameters as a function of the
baseline orientation.

The $\kappa$ CMa circumstellar disk seems to be elongated along B$_1$
but since we only have 3 visibility measurements  we cannot accurately
determine the angular position of the major-axis assuming an
elliptical circumstellar disk. The envelope flattening given by the
semi-major and semi-minor axis ratio, is about 2$\pm$0.7. Assuming
that the disk is geometrically thin (i.e. its opening angle is only a
few degree) we can estimate the range for the inclination angle i:
39$\degr$$<$i$<$68$\degr$. The lower limit of 39$\degr$ relies on the
lack of constraint on the disk opening angle. 

{\begin{table}[htbp]
    {\centering \begin{tabular}{c|ccc} \hline
        Base n$^o$ & 1 & 2 & 3 \\
        \hline
        Baseline & UT2-3 & UT3-4 & UT2-4 \\
        Length (m)& 42.7 &  59.3 & 80.8 \\
        P.A. ($\degr$) & 51.6 & 128 & 97.1 \\
        \hline
        V$_c$ &$>$0.93 &$>$0.93&$>$0.93\\
        V$_r$/V$_c$ &0.93$\pm$0.02 &0.95$\pm$0.02&0.9$\pm$0.02\\
        V$_r$ &0.85$<$V$_r$$<$0.95 &0.87$<$V$_r$$<$0.97&0.82$<$V$_r$$<$0.92\\
        \hline
        V$_{ec}$ &$>$0.86 &$>$0.86&$>$0.86\\
        V$_{er}$ &0.69$<$V$_{er}$$<$0.85&0.75$<$V$_{er}$$<$0.91&0.60$<$V$_{er}$$<$0.76\\
        \hline
        $\phi$$_{ec}$ (mas)&$<$3.6 &$<$2.6&$<$1.9\\
        $\phi$$_{er}$ (mas)&3.7$<$$\phi$$_{er}$$<$5.5&2.0$<$$\phi$$_{er}$$<$3.6&2.6$<$$\phi$$_{er}$$<$3.4\\
        \hline
        $\phi$$_{ec}$ (D$_\star$)&$<$15.5&$<$11.2&$<$8.2\\
        $\phi$$_{er}$ (D$_\star$)&15.9$<$$\phi$$_{er}$$<$23.7&8.6$<$$\phi$$_{er}$$<$15.4&$11.2<$$\phi$$_{er}$$<$14.6\\
        \hline
      \end{tabular}\par}
    \caption{
      \footnotesize{
        Br$\gamma$ Visibilities measured in the continuum (V$_c$)
        and visibility drop within the Br$\gamma$ line
        (V$_r$/V$_c$). V$_r$ calculated from the measured V$_c$ and
        V$_r$/V$_c$ ratio. Deduced envelope contribution in the
        continuum (V$_{ec}$) and in the line (V$_{er}$) for each
        baseline. The corresponding angular diameters in the Br$\gamma$
        line ($\phi_{er}$) and the nearby continuum ($\phi_{ec}$) are
        computed using a uniform disk model for each envelope
        measurement. The corresponding extension in stellar radii are
        also given, assuming a 6R$_\odot$ star at 230pc.
      }
    }
    \label{table_visi}
  \end{table}}

\section{SIMECA modeling}
\begin{figure}[htbp]
  \begin{center}
    \includegraphics[height=3.5cm, width=7.0cm]{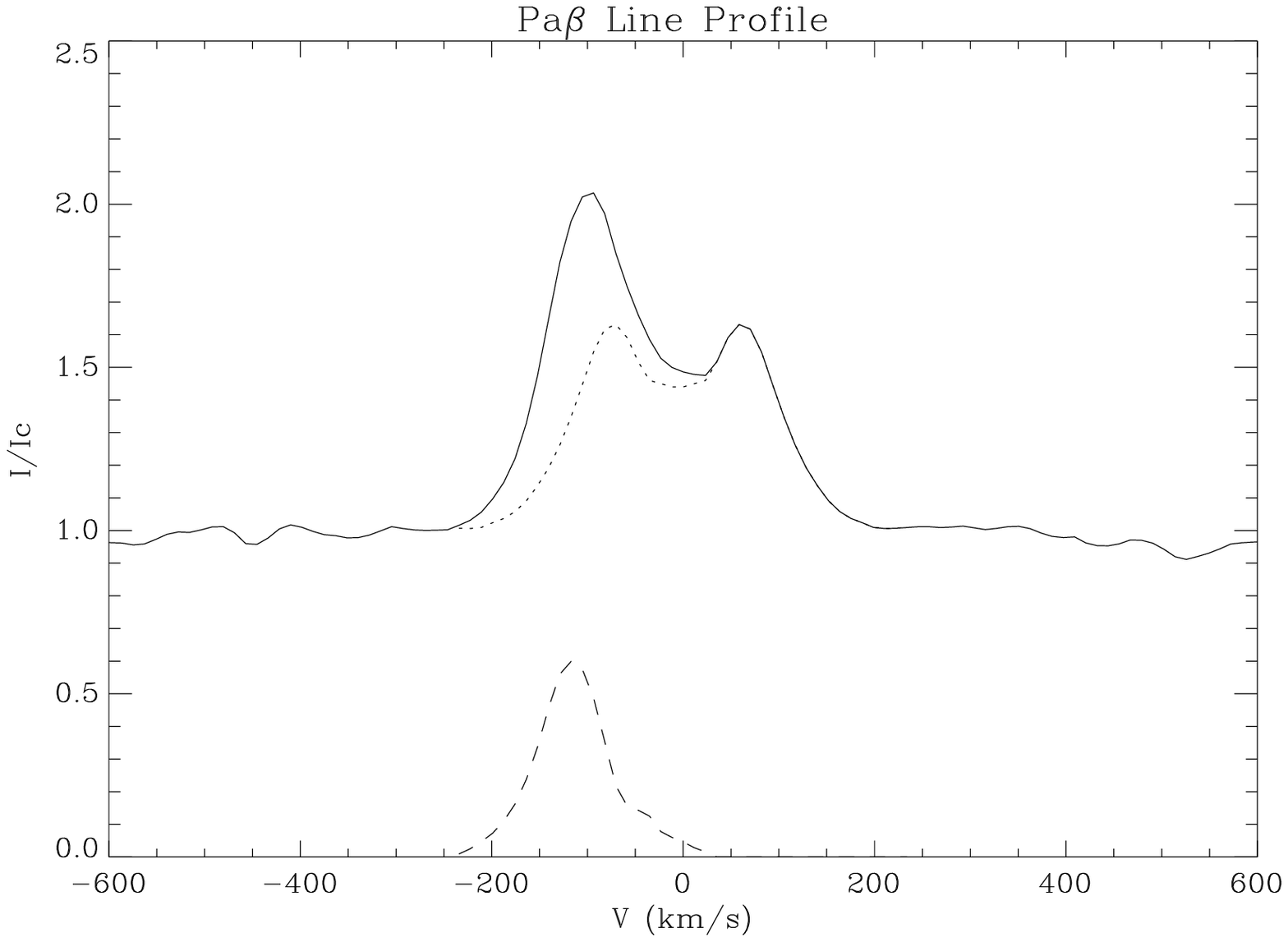}
    \includegraphics[height=3.5cm, width=7.3cm]{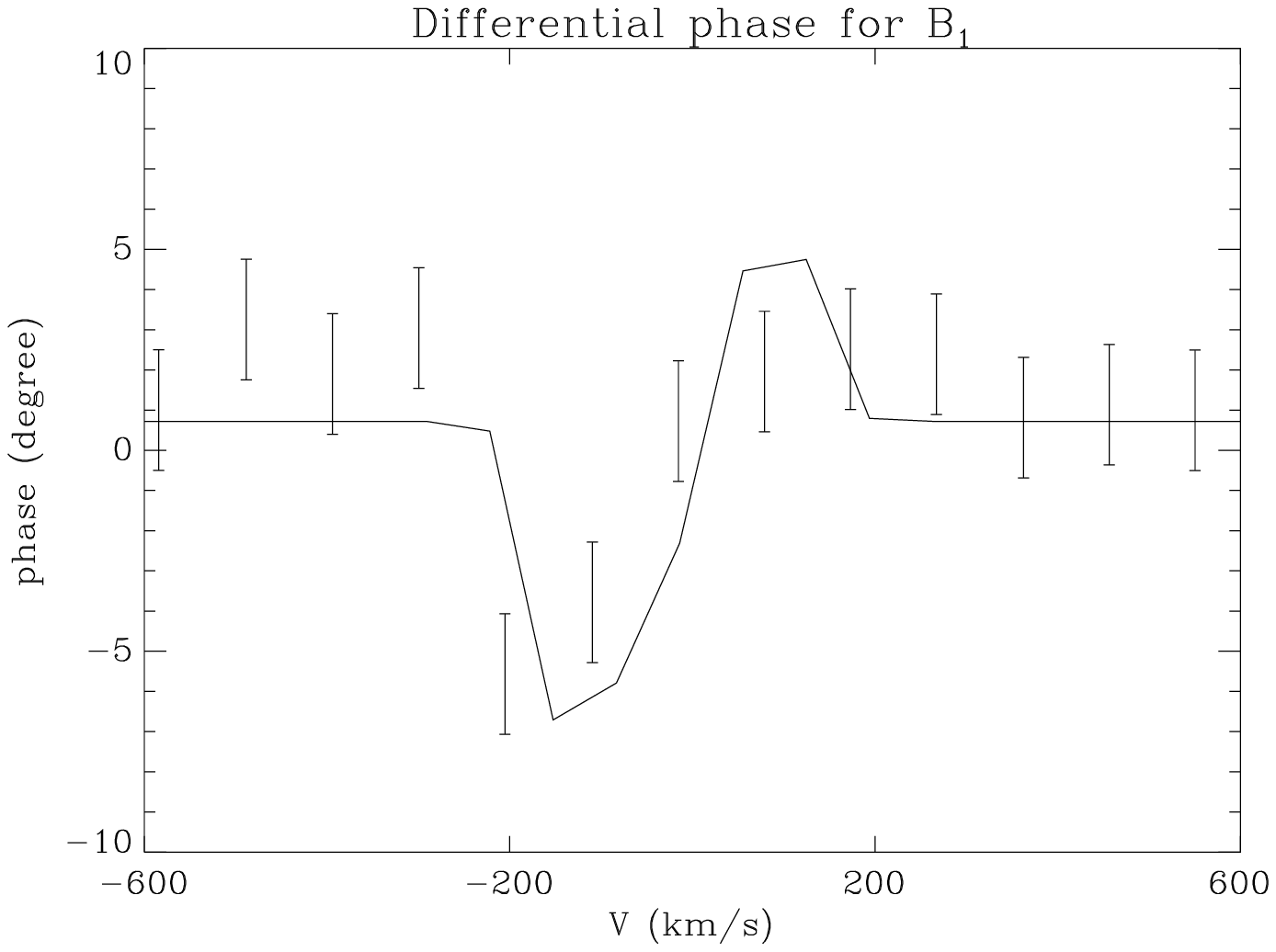}
    \caption{
      \footnotesize{
        Upper picture: $\kappa$ CMa Pa$\beta$ line profile
        observed in December 2005 at the Observatorio do Picos dos Dias,
        Brazil (solid line). Estimated symmetric part of the Pa$\beta$
        profile (dotted line) using an axi-symmetric model. The
        asymmetric residue corresponds to the emission of "one-armed"
        over-density (dashed line). Bottom picture: Differential phase
        variation measured along the B$_1$ baseline (dots with errors
        bars) and theoretical phase from the SIMECA code.
      }
    }
    \label{figure_pabeta}	
  \end{center}
\end{figure}

In order to obtain quantitative fundamental parameters of the central
star and its circumstellar disk, we used the SIMECA code developed by
Stee~\cite{Stee0} and Stee $\&$ Bittar~\cite{Stee3} to model the
$\kappa$ CMa circumstellar environment. Since this code was
axi-symmetric we made substantial modifications in order to introduce
a longitudinal dependence of the envelope density as evidenced from
the AMBER data plotted Fig.~\ref{figure_asymmodel}. To constrain the kinematics within the
disk we use a Pa$\beta$ line profile obtained in December 2005 at the
Observatorio do Pico do Dios, Brazil and plotted in Fig.~\ref{figure_pabeta}. This
profile is strongly asymmetric with a V/R double peak $\sim$1.3. This
V/R $>$ 1 is usually interpreted in terms of a viscous disk similar to
accretion disks where the gas and angular momentum are diffused
outward by magnetohydrodynamic viscosity (Lee et
al.~\citealp{lee}). Considering the time-dependent structure of
isothermal viscous disk, Okazaki~\cite{okazaki} showed that
"one-armed" density waves can propagate within the disk and should
reproduce the observed V/R variations from V/R$>$1 to V/R$<$1 seen in
the line profiles (Hummel \& Hanuschik \citealp{hummel}). Such
variations were detected for many Be stars, with period from a few
years to over a decade (Hanuschik et al.\citealp{hanuschik}; Telting
et al.~\citealp{Telting}). But in the case of $\kappa$ CMa the V/R
ratio has remained constant for the last twenty years (Dachs et
al.~\citealp{dachs}; Slettebak~\citealp{Slettebak}). 

In figure~5 we over-plotted the supposed "symmetric part" of the
Pa$\beta$ line profile, using an axi-symmetric model, and the
asymmetric residual that may be produced within the "one-armed"
oscillation over-density. This effect must be compatible with the
asymmetric differential phase variation across the Br$\gamma$ line for
the B$_1$ baseline plotted in the bottom part of Fig.~\ref{figure_pabeta}  since the
emitting regions in Pa$\beta$ and Br$\gamma$ must be very close each
together. The asymmetric contribution to the Br$\gamma$ emission is
about 20 to 30$\%$ of the total emission in this line whereas the mean
projected velocity of the inhomogeneity is -130$\pm$20 kms$^{-1}$.
Using a SIMECA model at 230pc we determined that the projected
separation between this over-density photocenter and the central star
is about 6.5$_\star$.\\

\begin{figure}[htbp]
  \begin{center}
    \includegraphics[width=8.0cm]{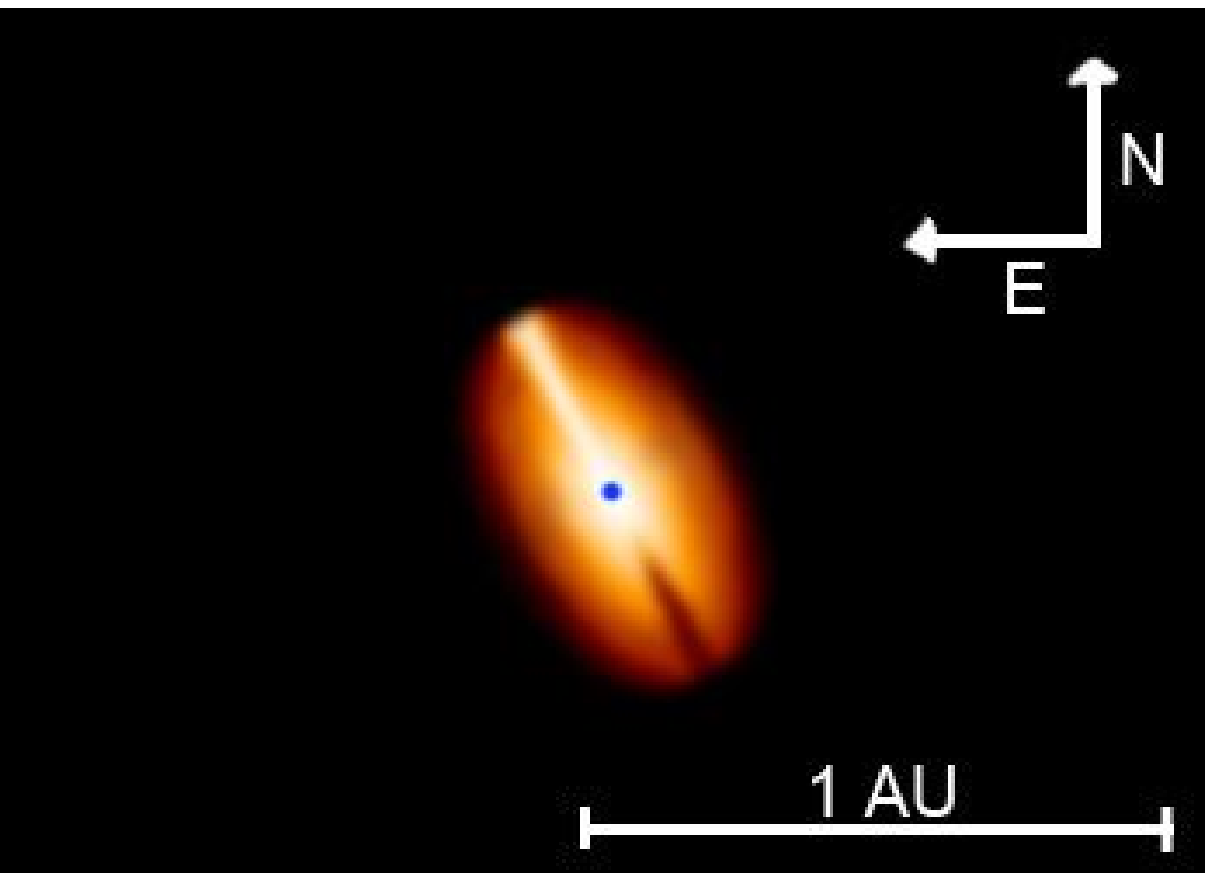}       
    \caption{
      \footnotesize{
        Intensity map in the continuum at 2.15 $\mu$m obtained
        with SIMECA for our best model parameters. The inclination angle
        is 60$\degr$, the central black dot represents the $\kappa$ CMa
        photosphere (0.25 mas), the bright part  in the equatorial disk
        is produced by the over-density which is oriented along the
        B$_{1}$ baseline. This over-density is also responsible for a
        30\% emission excess in the asymmetric  V part of the Br$\gamma$
        line.
      }
    }
    \label{figure_map_continu}
  \end{center}
\end{figure}

The parameters obtained  for our best model are given in Table
\ref{table_symparam} with the corresponding spectroscopic and
interferometric observables of Fig.~\ref{figure_asymmodel}. This best model
includes an over-density along the disk major axis at +20$\degr$,
corresponding to an over-luminosity of 30\% of the total flux in the
line, and the agreement with the VLTI/AMBER data, the SED
(Fig.~\ref{figure_sed}) and the Pa$\beta$ line profile is very good as it can
be seen Fig.~\ref{figure_asymmodel}. The very nice agreement with the
differential visibility and phase across the Br$\gamma$ line for the three
bases validates the chosen disk geometry and kinematics. The 2.1$\mu$m
continuum visibilities obtained with the 3 baselines, respectively
V$_1$=0.92, V$_2$=0.96 and V$_3$=0.94 are also compatible with the
0.93 lower limit measured with AMBER. The corresponding continuum
intensity map in the continuum at 2.15 $\mu$m is plotted
Fig.~\ref{figure_map_continu}. The evaluation on the uncertainties of the
parameters of our model is far below the scope of this letter and will
be studied in depht as soon as we get more constraining data.

\begin{table}
  \begin{center}
    \begin{tabular}{cc} \hline
      parameter    & value \\
      \hline
      $T_{\rm eff}$& 22\,500\,K $\pm$ 1000\\
      Radius& 6 R\( _{\sun } \) $\pm$ 1\\
      Inclination angle i & 60$\degr$ $\pm$ 10\\
      Equatorial rotation velocity & 240 km s\( ^{-1} \)  $\pm$ 20\\
      rotation law exponent & 0.32 $\pm$ 0.1\\
      Photospheric density ($\rho_{phot}$)&4. 10\( ^{-11} \)g cm\( ^{-3} \) $\pm$ 2 10\( ^{-11} \)\\
      Equatorial terminal velocity & 1 km s\( ^{-1} \) $\pm$ 10\\
      Polar terminal velocity & 1000 km s\( ^{-1} \) $\pm$ 100\\
      Polar mass flux & 2  10\( ^{-11} \)M\( _{\sun } \) year\( ^{-1} \) sr\( ^{-1} \) $\pm$ 0.5 10\( ^{-11} \)\\
      m1 & 10 $\pm$ 5\\
      m2 & 10 $\pm$ 2\\
      C1 & 30 $\pm$ 10\\
      Envelope outer radius & 23 R$_\star$ $\pm$ 2\\
      Major axis position & +28$\degr$ $\pm$ 5\\
      Over-density position & along the disk major axis\\ 
      \hline
    \end{tabular}
  \end{center}
  \caption{
    \footnotesize{
      Parameters for the $\kappa$ CMa central star and its
      circumstellar environment for the best axi-symetric model
    }
  }
  \label{table_symparam}
\end{table}

\section{Discussion}
Following recent VLTI/AMBER and VLTI/MIDI observations of  $\alpha$
Arae Meilland et al. \cite{meillanda} concluded that this classical Be
star fits very well within the classical scenario for the "Be
phenomenon", i.e. a fast rotating B star close to its breakup velocity
surrounded by a Keplerian circumstellar disk with an enhanced polar
wind.  This scenario was also confirmed for  the Be star Achernar by
Kervella \& Domiciano de Souza~\cite{kervella} using VLTI/VINCI data,
even if, for this latter case, the star was not in its active Be
phase, i.e without any strong emission line and no circumstellar
disk. Nevertheless, Achernar was still a nearly critical rotator and
was still exhibiting an enhanced polar stellar wind. We will see in the following
that $\kappa$ CMa does not fit very well within this classical scenario.

\subsection{Is $\kappa$ CMa a critical rotator ?}
If  $\kappa$ CMa was rotating close to its  breakup velocity,
  i.e. V$_c$=463kms$^{-1}$, the inclination angle would be around
  28$\degr$ in order to obtain a measured vsini =220 kms$^{-1}$). 
  With this inclination angle the maximum flattening
  corresponding to a geometrically very thin disk is 1.12. Since we
  measure a flattening of about 2$\pm$0.7 this inclination angle can
  be ruled out. In our best SIMECA model the star is rotating at only
  52$\%$ of its critical velocity. We may argue that the measured
    elongation is not the envelope major axis but rather the enhanced
    polar wind. In this case the projected axis of the Be envelope is
    not identical to the rotation axis of the star. Nevertheless, in
    order to obtain an asymmetry in the jet we need an extended
    optically thick disk, perpendicular to the jets directions, that
    may screen a least one part of the jet-like structure. Such
    extended optically thick disk should have been detected with the
    AMBER instrument which is not the case in our data.\\
    
     \noindent The value of the projected rotational velocity for an early-B star can be
    systematically affected by pseudo-photosphere, unrecognized
    optically thick parts of the Be envelope as shown by
    Harmanec~\cite{harmanec} for $\gamma$ Cas. He obtains for this star
    a vsini of 380 kms$^{-1}$ instead of the often quoted value
    of 230 kms$^{-1}$ from Sletteback~\cite{Slettebak}. Nevertheless,
    taking the largest value for $\kappa$ CMa from the literature from
    Zorec~\cite{zorec} who found a vsini=243kms$^{-1}$ we still
    obtain an inclination angle of 32$\degr$ which again is not in
    agreement with our measured flattening. Of course, if the
    discrepancy between the measured vsini and the "real" one is
    larger it may be possible that $\kappa$ CMa is still a critical
    rotator but it requires a factor of 2 between the measured and the
    true vsini which we found unrealistic.  Even if Townsend et
    al.~\cite{townsend} include the gravity darkening effect on the vsini
    values of rigid early-type rotators, assuming
    a rotation rate $\Omega$/$\Omega_{c}$ of 0.95, they conclude that
    classic vsini determinations for B0 to B9-type stars can be
    underestimated by 12 to 33\%, far from a factor of 2. Moreover, a recent
    paper by Fr\'emat et al.\cite{fremat} studying the effect of the
    gravitational darkening on the determination of fundamental
    parameters in fast rotating B-type stars found that on average the
    rate of angular velocity of Be stars attains only
    $\Omega$/$\Omega_{c}$$\sim$0.88.\\
    
    \noindent In this last paper, Fr\'emat et
    al.\cite{fremat} estimate $\kappa$ CMa's effective temperature to
    be 25790 $\pm$ 713 K, a value significantly larger than the 22500
    K used in our modeling. Moreover, Harmanec \cite{harmanec0} found
    a positive correlation between the emission strength and
    brightness in the optical. Therefore we may use the minimum
    observed V magnitude of about 3.5 to estimate the radius of the
    central star. Combining with the Hipparcos parallax and its error
    we obtain a radius between 9 and 14 solar radii. Using the
    T$_{eff}$ of 25790 K and a radius of 14 R$_{\sun}$ we obtain a
    stellar luminosity larger by a factor 8 compared to our modeling
    and thus it is not possible to obtain a good agreement with the
    observed SED plotted Fig.\ref{figure_sed}. Thus we are more confident
    with our 6 R$_{\sun}$ used for our modeling and our finding that
    $\kappa$ CMa seems not to be a critical rotator. Nevertheless,
    regarding the uncertainties and the large errors of all
    measurements the breakup velocity cannot be totally excluded.

\subsection{Is the rotation law within the disk Keplerian ?}
   A Keplerian rotating law would produce a narrower
   double-peak separation in the Pa$\beta$ line profile. Using a
   simple axi-symmetric Keplerian disk model the double-peak
   separation would be about 90kms$^{-1}$ whereas we measure an
   asymmetric double-peak separation of about 160kms$^{-1}$.  Even if
   we subtract the emission of the over-density producing a larger
   double-peak separation by contributing to the V peak of the
   emitting Pa$\beta$ line, we still obtain a double-peak separation
   of about 120kms$^{-1}$ (see Fig.~\ref{figure_pabeta}).  The exponent of the rotation
   law used for our best SIMECA model is 0.32 whereas it should be 0.5
   for a purely Keplerian disk.\\
   
   \noindent We may argue that Be stars vary
     strongly in time and thus line profiles shapes are time
     dependent. For instance actual H$\alpha$ line profiles show a
     strong emission with a single peak whereas Bahng \&
     Hendry~\cite{bahng} saw a double-peaked H$\alpha$ emission line,
     with the same double-peak separation of 160 km$^{-1}$ we obtained
     for Pa$\beta$, with a shell core on their high-dispersion
     spectra. Nevertheless, these line variations are certainly related
     to the formation and disappearance of the circumstellar disk
     around the star as shown by  Rivinius et al.~\cite{rivinius} and
     Meilland et al.~\cite{meillandb}. Whatever the model is, a
     double-peak line profile is a clear signature of an extended
     rotating disk, at least if the kinematics is not dominated by a
     strong stellar wind in the equatorial region as shown by Stee \&
     de Ara\`ujo~\cite{Stee0}. This double-peak separation is a good
     indication of the disk extension as shown by Huang~\cite{huang};
     Hirata \& Kogure~\cite{hirata} and Stee \& de
     Ara\`ujo~\cite{Stee0}. We measure v$_{disk}$ sini at the disk
     outer radius (R$_{disk}$) from the peaks separation, where
     v$_{env}$ is the rotational disk velocity at R$_{disk}$.  Thus we
     can write:

     \begin{equation}
       v_{disk} \sin(i) =v_{\star} \sin(i)
       \left(\frac{R_{disk}}{R_{\star}} \right)^{-\beta},
     \end{equation}

     where v$_{\star}$ is the star rotation at its photosphere.\\

\noindent Assuming a Keplerian rotation ($\beta$ =0.5) we obtain, using
     Eq. 3, R$_{disk}$=13.5 R$_{\star}$ which is about 2$\sigma$ from
     the 19.8 R$_{\star}$ interferometric measurement, assuming that
     the measured elongation is the envelope major axis and not an
     enhanced polar wind (see discussion in the previous point). Note
     that these 19.8 R$_{\star}$ found are obtained assuming a
     uniform disk for the envelope and thus is certainly a lower
     limit to the "true" disk extension in the Pa$\beta$ line. Thus it seems 
     difficult to maintain a Keplerian rotation within the disk of $\kappa$ CMa.

  \begin{figure}[htbp]
    \begin{center}
      \includegraphics[width=8.0cm]{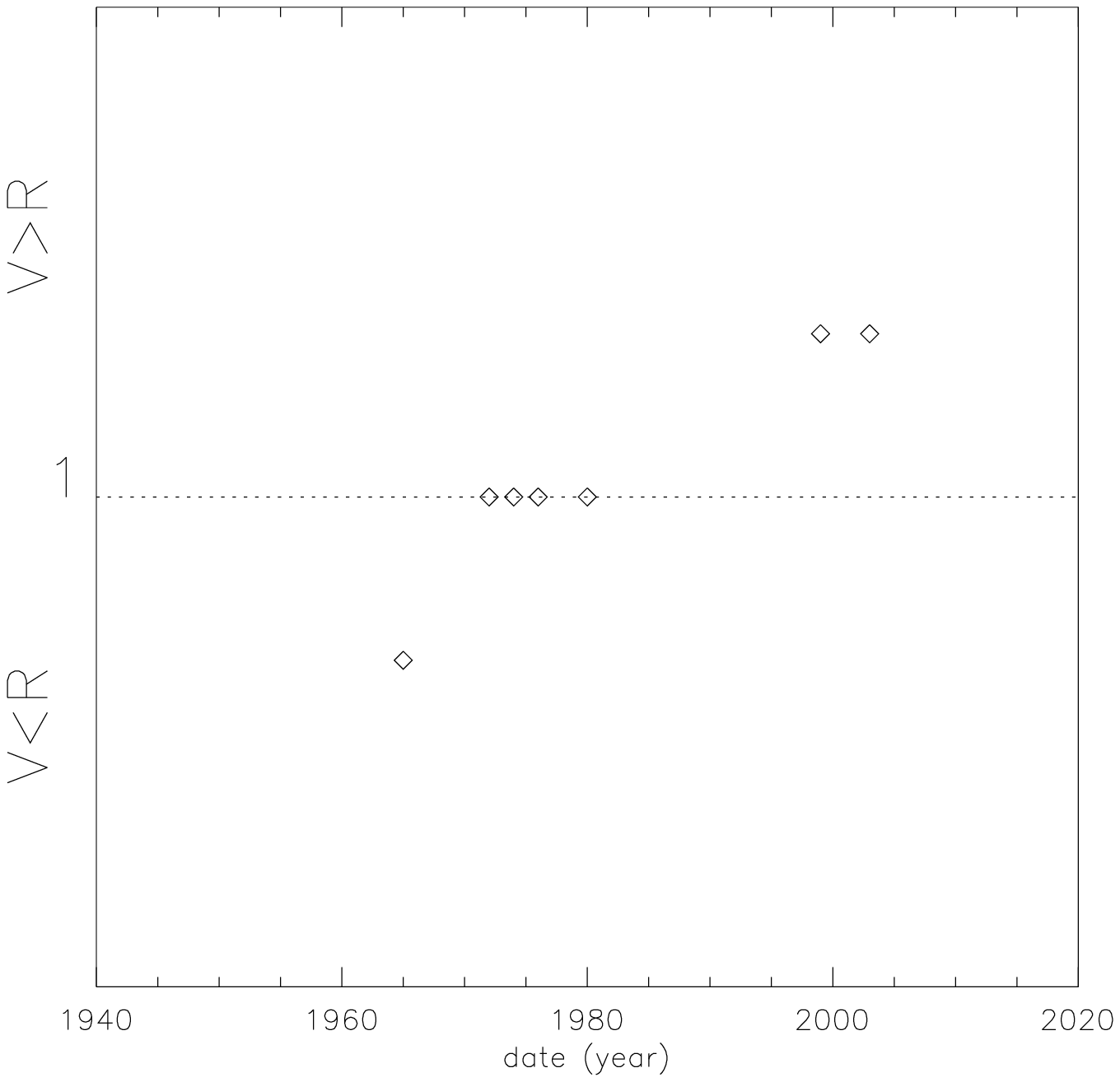}       
      \caption{V/R variations obtained from the literature between
        1965 and 2003, respectively from  Jaschek~\cite{Jaschek},
        Slettebak~\cite{Slettebak0}, Banerjee~\cite{banerjee} and this
        work.}
      \label{figure_VR_variations}
    \end{center}
  \end{figure}

\subsection{Is the "one-armed" viscous disk model a possible scenario for $\kappa$ CMa  ?}
The asymmetry presently detected in the disk of $\kappa$ CMa seems to be hardly 
explained within the "one-armed" viscous disk framework. 
Following the viscous disk models by Okazaki~\cite{okazaki} and the observational detection of
  "one-armed" oscillations in the disk of $\zeta$ Tau by Vakili et
  al.~\cite{vakili} and $\gamma$ Cas by Berio et al.~\cite{berio}, the
  precessing period (P) of such oscillations should be confined within
  a few years up to about twenty years for the longer ones. On
    the other side, we tried to compile all the observational data
    available in order to obtain a "quasi-period" for the V/R
    variations. First we must mention that the V/R variations occur
    during the time intervals of observable presence of Be envelopes
    and that they can show long-term, medium-term as well as rapid
    changes (Dachs~\citealp{dachs1}). Moreover, the very strong
    H$\alpha$ line profile is not really suitable to V/R measurements
    since it is single-peaked and the illusion of apparent V/R changes
    can be related to the presence of telluric lines. Finally,
    compiling the data between 1965 and 2003 for $\kappa$ CMa
    respectively from Jaschek~\cite{Jaschek},
    Slettebak~\cite{Slettebak0}, Banerjee~\cite{banerjee} and this
    work, we were not able to deduce an estimation of a quasi-period 
    (Fig.~\ref{figure_VR_variations}). Several authors suggested a very
     long V/R variation (i.e. Okazaki P$>$28 years). This is a possibility but
an equally plausible possibility is that the star had two episodes
of V/R changes with much shorter cycle length separated by a period
of quiescence documented by Dachs et al.~\citealp{dachs}; Slettebak~\citealp{Slettebak}.
     More observations are needed 
     since, if this first possiblity could be confirmed, then there is a problem for the one-armed model.
     This "pseudo-period" would be too long compared to theoretical predictions which can be hardly 
      longer than two decades for a disk with a radius $\sim$ 23R$_{\star}$
      (Okazaki, private communication). Moreover, the fact that this 
      over-density remains confined along the major axis of the 
      disk seems to be a very nice coincidence...\\
      
 \noindent Clearly, further observations are mandatory to confirm these
  conclusions or to see if other physical phenomena occurred within the
  circumstellar disk of $\kappa$ CMa.


\begin{acknowledgements}
  We thanks Atsuo Okasaki for his useful comments about the viscous
  discs models. We acknowledge the remarks of the referee Petr
  Harmanec which help to greatly improve the paper. We thank David
  Chapeau and Damien Mattei for the SIMECA code developments support.

  The AMBER project has benefited from funding from the French Centre
  National de la Recherche Scientifique (CNRS) through the Institut
  National des Sciences de l'Univers (INSU) and its Programmes
  Nationaux (ASHRA, PNPS). The authors from the French laboratories
  would like to thank the successive directors of the INSU/CNRS
  directors.  The authors from the the Arcetri Observatory acknowledge
  partial support from MIUR grants and from INAF grants.
  C. Gil work was supported in part by the Funda\c{c}\~ao para a
  Ci\^encia e a Tecnologia through project POCTI/CTE-AST/55691/2004
  from POCTI, with funds from the European program FEDER.

  This research has also made use of the ASPRO observation preparation
  tool from the JMMC in France, the SIMBAD database at CDS, Strasbourg
  (France) and the Smithsonian/NASA Astrophysics Data System
  (ADS). This publication makes use of data products from the Two
  Micron All Sky Survey.

  The AMBER data reduction software \texttt{amdlib} has been linked with the open source software Yorick\footnote{\texttt{http://yorick.sourceforge.net}} to provide the user friendly interface \texttt{ammyorick}. They are freely available on the AMBER website \texttt{http://amber.obs.ujf-grenoble.fr}.
\end{acknowledgements}

\end{document}